\documentclass[sigconf]{acmart}
\usepackage{svg}

\usepackage{xcolor}
\usepackage{calc}

\usepackage{geometry}
\usepackage{longtable}
\usepackage{array}
\usepackage{xcolor}
\usepackage{booktabs}
\usepackage{enumitem}
\usepackage{wrapfig}
\usepackage{todonotes}

\usepackage[utf8]{inputenc}
\usepackage{textgreek}

\setcopyright{none}
\renewcommand\footnotetextcopyrightpermission[1]{}

\definecolor{YingqiangNote}{HTML}{E6A23C}

\newif\ifshowyingqiangnotes
\showyingqiangnotestrue 

\AtBeginDocument{%
  }

\setcopyright{acmlicensed}
\copyrightyear{2026}
\acmYear{2026}
\acmDOI{XXXXXXX.XXXXXXX}

\acmConference[CUI '26]{CUI '26: Extended Abstracts of the 2026 Conference on Conversational User Interfaces }{July 21st–24th, 2026}{Bremen, Germany}
\acmISBN{978-1-4503-XXXX-X/2018/06}

\begin{document}

\title{Demonstration of Adapt4Me: An Uncertainty-Aware Authoring Environment for Personalizing Automatic Speech Recognition to Non-normative Speech}

\author{Niclas Pokel}
\email{npokel@ethz.ch}
\orcid{0009-0005-6429-2019}
\affiliation{%
  \institution{Institute of Neuroinformatics, University of Zurich and ETH Zurich}
  \city{Zurich}
  \country{Switzerland}
}

\author{Yiming Zhao}
\email{yimizhao@student.ethz.ch}
\affiliation{%
  \institution{Department of Computer Science, ETH Zurich}
  \city{Zurich}
  \country{Switzerland}
}

\author{Pehuén Moure}
\email{pehuen@ini.ethz.ch}
\orcid{0009-0001-1631-8238}
\affiliation{%
  \institution{Institute of Neuroinformatics, University of Zurich and ETH Zurich}
  \city{Zurich}
  \country{Switzerland}
}

\author{Yingqiang Gao}
\authornote{Corresponding authors.}
\email{yingqiang.gao@cl.uzh.ch}
\orcid{0009-0000-0876-621X}
\affiliation{%
  \institution{Department of Computational Linguistics, University of Zurich}
  \city{Zurich}
  \country{Switzerland}
}

\author{Roman Boehringer}
\authornotemark[1]
\email{romaboeh@ethz.ch}
\orcid{0000-0003-2856-3262}
\affiliation{%
  \institution{Institute of Neuroinformatics, University of Zurich and ETH Zurich}
  \city{Zurich}
  \country{Switzerland}
}

\renewcommand{\shortauthors}{Pokel et al.}

\begin{abstract}

 Personalizing Automatic Speech Recognition (ASR) for non-normative speech remains challenging because data collection is labor-intensive and model training is technically complex. To address these limitations, we propose \textit{Adapt4Me}, a web-based decentralized environment that operationalizes Bayesian active learning to enable end-to-end personalization without expert supervision. The app exposes data selection, adaptation, and validation to lay users through a three-stage human-in-the-loop workflow: (1) rapid profiling via greedy phoneme sampling to capture speaker-specific acoustics; (2) backend personalization using Variational Inference Low-Rank Adaptation (VI-LoRA) to enable fast, incremental updates; and (3) continuous improvement, where users guide model refinement by resolving visualized model uncertainty via low-friction top-$k$ corrections. By making  epistemic uncertainty explicit, \textit{Adapt4Me} reframes data efficiency as an interactive design feature rather than a purely algorithmic concern. We show how this enables users to personalize robust ASR models, transforming them from passive data sources into active authors of their own assistive technology.

\end{abstract}




\begin{teaserfigure}
  \centering
  \includegraphics[width=0.9\textwidth]{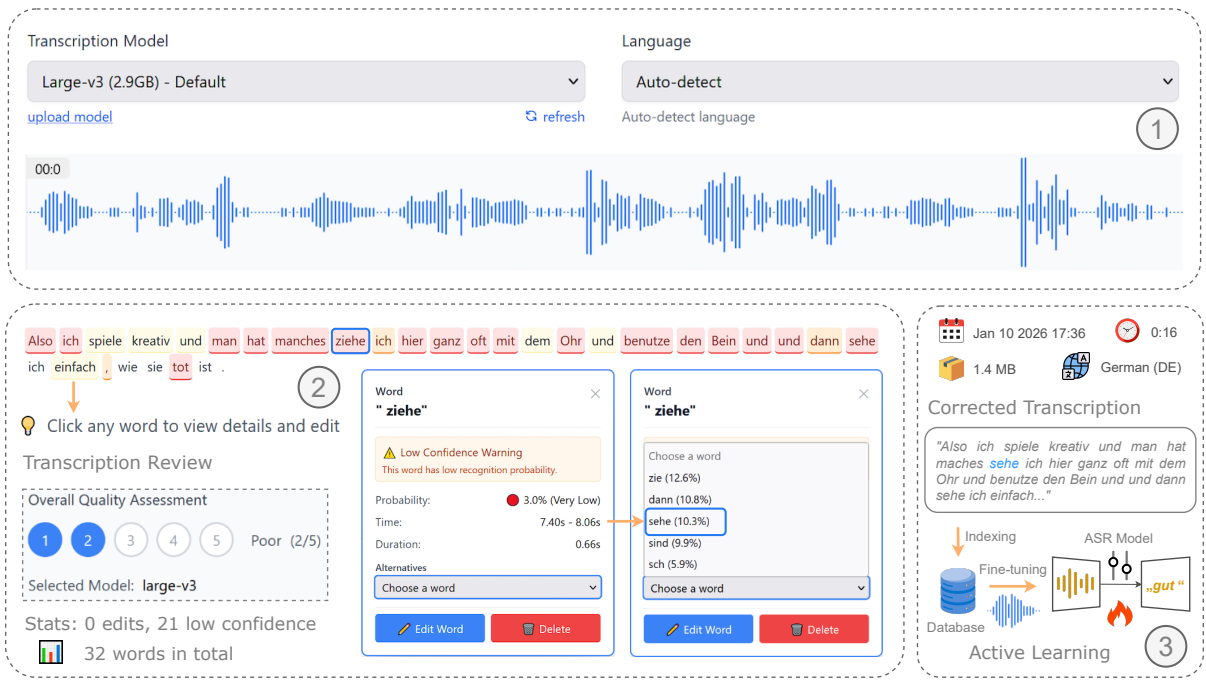} 
  \caption{\textbf{\textit{Adapt4Me}: an end-to-end, human-in-the-loop ASR personalization system for non-normative speech}. \textcircled{1} The system prompts the user with high-information sentences for initial recordings; \textcircled{2} Word-level model uncertainty is visualized through color-coding, guiding users directly to likely identify and correct transcribing errors; \textcircled{3} Users resolve errors via low-effort top-$k$ best corrections powered by a Bayesian backend, with validated edits continuously fed back into active learning and model fine-tuning.}
  \Description{A three-panel screenshot of the web application. The left panel shows a microphone icon and a prompt. The center panel shows transcribed text with the word ``Pellikan'' highlighted in red. The right panel shows a dropdown menu suggesting 'Pelican' as a correction. }
  \label{fig:teaser}
\end{teaserfigure}


\maketitle

\pagestyle{plain}

\section{Introduction}

State-of-the-art ASR models like Whisper \cite{radford2023robust} or Wav2Vec \cite{schneider2019wav2vec} have achieved remarkable performance on normative speech, yet they exclude individuals with motor speech disorders (e.g., dysarthria) \cite{maassen2007speech, Duffy2005MotorSpeech, ballati2018assessing} or structural impairments of the speech apparatus (e.g., Apert syndrome) \cite{Zimmerman2020}. While previous work has explored fine-tuning foundation ASR models for non-normative speech, error rates remain substantially higher than those observed for normative speech \cite{Singh2024, KimSung2023, liu2022recent}. 
This lack of generalization to non-normative speech further reinforces the marginalization of individuals with speech impairments, as their specific communicative needs are rarely addressed by modern speech technologies, thereby limiting their access to the benefits of AI.

Nevertheless, although personalizing ASR models for user-specific transcription is technically feasible, it remains largely inaccessible in practice. Conventional model personalization typically requires hours of voice recordings, which is often exhaustive for individuals with speech impairments and places a substantial burden on family members and caregivers \cite{cronin2020academic, hitchcock2015social, page2022communicative, sarsenbayeva2022methodological}.
This process becomes even more burdensome when personalized models must be repeatedly optimized with newly recorded data to adapt to the evolving acoustic characteristics of individuals with speech impairments, who are often everyday ASR users without expert knowledge.

Users often experience ASR systems as “black-box” transcribers, with little insight into which phonemes cause recognition errors\citep{kuhn2025evaluating, glazer2025beyond}. This opacity is largely due to the absence of token-level uncertainty visualizations that reveal the system’s transcription difficulties, preventing users from providing targeted and effective feedback in subsequent fine-tuning iterations. Previous work in HCI suggests that well-designed uncertainty visualizations enable users to make better-informed decisions and act as effective supervisors of imperfect AI systems \cite{10.1145/2858036.2858558}.

We argue that effective ASR personalization requires a shift from passive data collection to interactive curating. This shift hinges on exposing the model’s epistemic uncertainty, its awareness of what it does not yet model reliably, so users can guide personalization in a principled and efficient way.  Recent work in Interactive Machine Learning has shown that users value transparency and agency when training assistive tools \cite{goodman2025spectra, kacorri2017people}. This shift is particularly critical for implementation of ASR personalization in real-world settings, where minimizing user effort is essential for enabling continuous adaptation to newly recorded speech data, but also for ensuring sustained user engagement.

We present \textit{Adapt4Me}, a web-based, accessible, and low-effort environment that operationalizes Bayesian Active Learning (AL) for personalizing ASR for non-normative speech. By integrating an uncertainty-aware VI-LoRA-based ASR model \citep{pokel2025vilora} into a user-friendly interface, \textit{Adapt4Me} has the following advantages over existing work \citep{tobin2024automatic, martin2025project}:
\begin{itemize}[left=0pt]
    \item \textbf{Efficient ``Cold Start''}: A profiling method using Greedy Biphone Coverage \cite{pokel2025adaptingfoundationspeechrecognition} that bootstraps personalization in minutes of training data rather than hours;
    \item \textbf{Uncertainty-Guided Feedback}: A visual interface that highlights confused phonemes, guiding users to provide personalization data for active learning through direct human-computer interaction;
    \item \textbf{Longitudinal Resilience}: A lifecycle management feature handling non-linear speech changes (e.g., post-surgery or puberty voice change).
\end{itemize}

\textit{Adapt4Me} is a proof-of-concept prototype that demonstrates how advances in ASR models can be converted into tangible solutions to improve communication accessibility for individuals with speech impairments. By bridging the gap between technological advances and real-world deployment, \textit{Adapt4Me} enables efficient and personalized ASR adaptation that can support speech-impaired users over the long term.

\section{Motivation}



\begin{table}[!htb]
    \centering
    \caption{Comparison of hours of existing speech datasets (recording hours are reduced by redundant microphones).}
    \label{tab:datasets}
    \begin{tabular}{llr}
        \toprule
        \textbf{Domain} & \textbf{Dataset} & \textbf{Hours} \\
        \midrule
        Normative & Common Voice (EN) \cite{ardila-etal-2020-common} & $\sim$3,800 \\
        Normative & LibriSpeech (EN) \cite{7178964} & $\sim$1,000 \\
        \midrule
        Dysarthric & SAP (EN) \cite{SAPdataset} & $\sim$415 \\
        Dysarthric & UA-Speech (EN) \cite{kim2008dysarthric} & $\sim$9 \\
        Dysarthric & BF-Sprache (DE) \cite{pokel2025adaptingfoundationspeechrecognition} & $\sim$2.5 \\
        Dysarthric & TORGO (EN) \cite{Rudzicz2012} & $\sim$3 \\
        \bottomrule
    \end{tabular}
\end{table}
The rich diversity of speech impairments cannot be addressed simply by collecting more data, as variability across individuals is substantially higher than in normative speech populations \cite{xiong2019phonetic, troger2024automatic, van2023automatic}. This high inter-speaker variance poses significant challenges for training personalized ASR models. As shown in Table~\ref{tab:datasets}, available non-normative speech resources are orders of magnitude smaller than their normative counterparts. Consequently, this data-scarcity bottleneck limits the applicability of traditional personalization approaches, which are typically data-intensive and depend on extensive expert annotation, making them impractical for deployment in a  home settings \cite{10.1145/3442188.3445870}. 
In addition, standard personalization approaches typically fine-tune models on small amounts of user data, making them prone to overfitting and catastrophic forgetting \cite{MCCLOSKEY1989109, FRENCH1999128}. While Parameter-Efficient Fine-Tuning (PEFT \cite{he2022towards}) methods such as LoRA \cite{hu2022lora} partially alleviate these issues, they do not address a more fundamental problem: the lack of an effective and accessible workflow for personalizing ASR models. Without human-in-the-loop (HITL \cite{wu2022survey, mosqueira2023human}) guidance, users often expend effort recording redundant data, namely utterances that the model already transcribes well, while failing to gather data that is most informative for further reducing recognition errors. This mismatch between user effort and model improvement underscores the necessity of systems that enable data-efficient personalization while minimizing cognitive and physical burden on end users.


To address both data scarcity and the lack of effective personalization workflows in ASR, \textit{Adapt4Me} is designed following the principle of active learning \cite{hakkani2002active, settles2009active}. Rather than treating users as passive data providers, \textit{Adapt4Me} positions them as active participants in the personalization loop, a paradigm aligned with interactive machine learning principles \cite{PowertothePeople}. In this setting, users inspect transcription outputs, judge their quality, and correct erroneous predictions to directly guide model adaptation. This HITL process enables the system to focus data collection on informative samples that contribute most to performance improvement. Through \textit{Adapt4Me}, we demonstrate that effective ASR personalization systems must go beyond algorithmic adaptation alone: they must reduce human effort through targeted interaction and provide actionable feedback that supports iterative model improvement in real-world, home-based settings.


\section{System Architecture}

\begin{figure*}[t]
    \centering
    \includegraphics[width=0.9\textwidth]{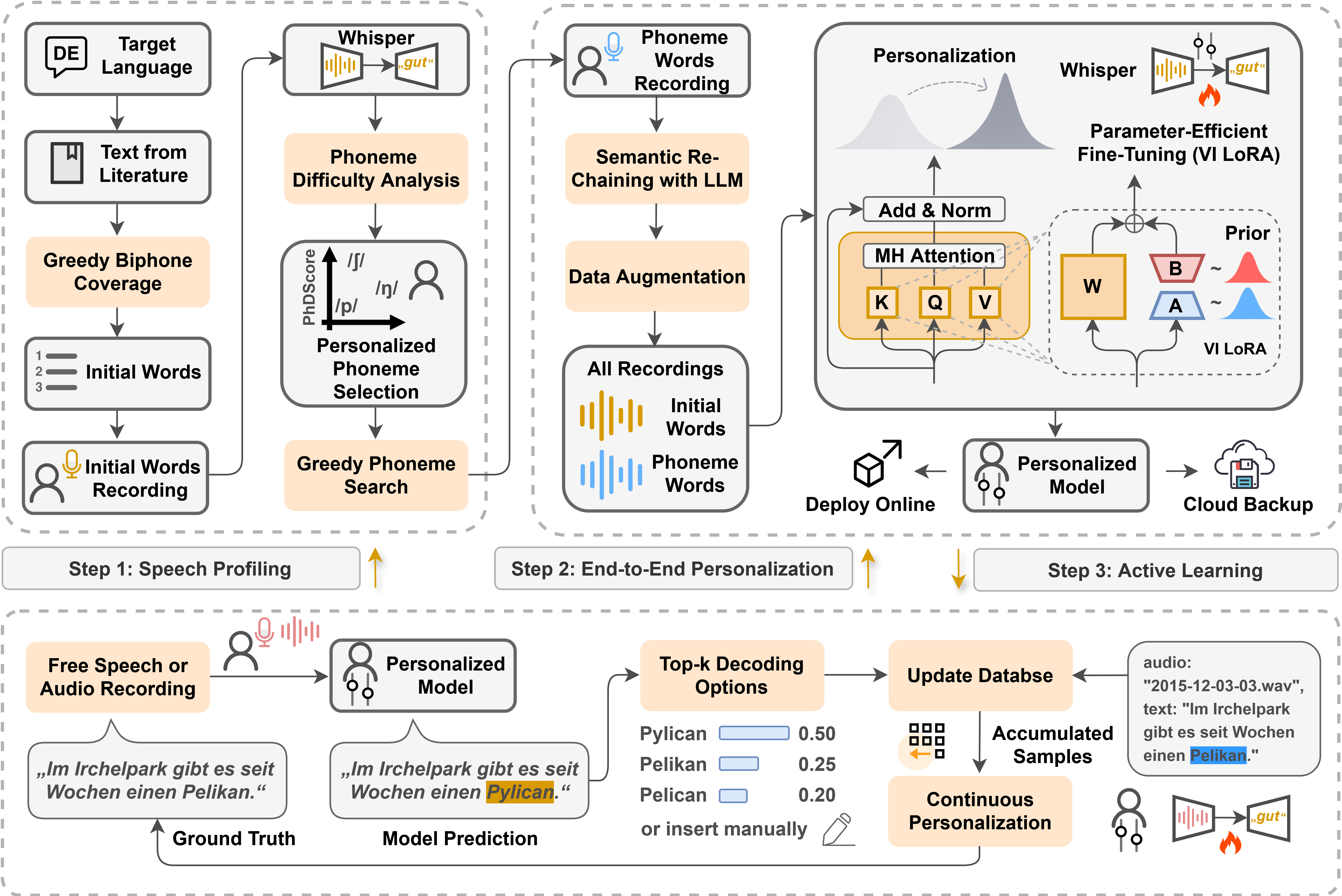}
    \caption{\textbf{The Adapt4Me Personalization Lifecycle.} 
    \textbf{Step 1 (Speech Recording):} The system creates a "Cold Start" profile using a minimized set of 500 words selected via Greedy Biphone Coverage and LLM Prompting \cite{pokel2025adaptingfoundationspeechrecognition}.
    \textbf{Step 2 (Model Personalization):} The backend adapts the model. It uses Semantic Re-chaining (SRC) \cite{pokel2025adaptingfoundationspeechrecognition} to generate context, calculates the Phoneme Difficulty Score (PhDScore) to search for highly informative training-samples \cite{pokel2025phonemes}, and fine-tunes the VI-LoRA adapters \cite{pokel2025vilora} (Top Right).
    \textbf{Step 3 (Active Learning):} The continuous interaction loop. The model transcribes free speech, highlights uncertain words (e.g., ``Pylikan''), and generates context-aware N-best suggestions to minimize annotation effort.}
    \label{fig:flow}
\end{figure*}

The architecture of \textit{Adapt4Me} follows the three-step workflow visualized in Figure \ref{fig:flow}, starting from speech recording to continuous, uncertainty-guided adaptation.

\textbf{Step 1: Speech Profiling.}
To bootstrap personalization for a new user, we employ a \textit{Greedy Biphone Coverage} algorithm \cite{pokel2025adaptingfoundationspeechrecognition}. The system first selects a minimal set of initial words from the literature to maximize phonetic variance. These words are used as seeds for an LLM prompt to generate semantically and structurally coherent training samples.

\textbf{Step 2: End-to-End Personalization.}
Once the initial audio is captured, the backend personalizes the model using VI-LoRA \cite{pokel2025vilora}. This parameter-efficient method stabilizes fine-tuning on small datasets and, crucially, quantifies \textit{epistemic uncertainty} \cite{pokel2025phonemes}. We aggregate this uncertainty to calculate a \textit{Phoneme Difficulty Score}, identifying the user's specific articulatory struggles \cite{pokel2025phonemes}. This diagnosis drives the \textit{Semantic Re-chaining} engine to synthesize new, targeted sentence prompts rich in these difficult phonemes, constructing a personalized training curriculum for the next phase.
    
\textbf{Step 3: Active Learning}.
Users correct the free speech predictions, guided by word-level uncertainty scores. A key challenge is that standard top-$k$ alternatives from ASR models often lack semantic coherence. To address this, we employ a two-pass decoding strategy: a \textit{coherent pass}, which produces semantically consistent transcriptions, and a \textit{variation pass}, which selectively re-samples only high-uncertainty words while keeping their surrounding context fixed. 


\section{Human-Computer Interaction Design}

To guide user correction, \textit{Adapt4Me} highlights low-confidence words in model transcriptions (e.g., ``Pylikan'' in Figure~\ref{fig:flow}) using entropy-based uncertainty scores. These uncertainty scores serve two complementary purposes: (1) as a \textit{speech diagnostic tool}, offering insights into how the model interprets user speech by revealing words or phonemes that consistently cause transcription errors; and (2) as a \textit{task management tool}, directing user attention to the most probable errors, thereby eliminating the need to proofread correctly transcribed segments and reducing cognitive load.

Motor impairments frequently co-occur with speech impairments in individuals with neurodevelopmental disorders \cite{lancioni2025people}, which can make fine-grained motor actions such as typing corrections difficult and fatiguing. As a result, manually editing ASR outputs can become a significant barrier for such users \cite{10.1145/1502650.1502685}. To address this challenge, \textit{Adapt4Me} replaces typing-based corrections with a context-aware top-$k$ selection mechanism, allowing users to correct errors by simply selecting the appropriate word from a short list of alternatives. Manual insertion remains available as a fallback. By transforming error correction from a typing task into a lightweight selection task, the system substantially reduces physical effort and improves accessibility for users with motor limitations.


To support lifelong ASR personalization, \textit{Adapt4Me} is designed to accommodate the dynamic nature of speech impairments. The system incrementally adapts to gradual changes, such as a child’s speech development, as users continuously contribute corrected samples. In contrast, major physiological changes, including post-surgical recovery \cite{Reddy2023} or puberty-related voice changes \cite{pinheiro2024neural}, may render an existing personalized acoustic model less effective. In such cases, users can re-initiate the system from the initial recording stage to establish a new acoustic baseline, while preserving previously learned semantic and lexical personalization. This design prevents a complete loss of progress and supports long-term, evolving use.

When transcribing with \textit{Adapt4Me}, ten forward passes produce an inference latency of $\approx 2$ s on a standard GPU (e.g. NVIDIA GeForce RTX 3090). This is acceptable for the home setting, as \textit{Adapt4Me} was designed primarily to personalize ASR models rather than the real-time user experience. 
We employ a client-server setup, with a webapplication (Client) running on edge devices like tablets or smartphones, familiar to children. The computationally intensive personalization is offloaded to a secure cloud backend. This ensures that families do not need high-end GPUs for home deployment.

Unlike sound-treated clinical environments, home settings are often acoustically noisy. \textit{Adapt4Me} therefore incorporates a lightweight Signal-to-Noise Ratio (SNR) check that runs locally in the browser to prevent the submission of low-quality recordings, such as speech contaminated by background television noise, that could degrade system performance. In addition, \textit{Adapt4Me} supports collaborative use by multiple family members. Prompts are presented in large, simple fonts for the speech-impaired child, while top-$k$ suggestions and uncertainty visualizations are positioned to enable parental oversight, facilitating shared interaction and supervision during the personalization process.

\section{Envisioned Conference Experience}
To effectively demonstrate this home-based, collaborative workflow in a conference setting, we will present \textit{Adapt4Me} via a local web interface deployed on a provided laptop and tablet. Because conference attendees typically possess normative speech, and because live model fine-tuning exceeds feasible booth dwell times, the demonstration is designed as an interactive, simulated authoring scenario. 

Attendees will step into the role of the ``human-in-the-loop'' (e.g., acting as the caregiver or user), tasked with curating pre-recorded, anonymized non-normative speech samples. This workflow allows attendees to directly experience the system's core HCI contributions: interpreting the token-level uncertainty visualizations and utilizing the low-friction top-$k$ selection mechanism to correct transcription errors without typing. 

Following the correction phase, attendees will experience a live comparison, contrasting the semantic hallucinations of a baseline Whisper model with the phonetically accurate output of an \textit{Adapt4Me}-personalized model. The physical setup natively supports the dyadic interaction described above, ensuring a fast, engaging 3-minute flow that accommodates high booth throughput while mitigating the challenges of the noisy conference exhibition hall.

\section{Evaluation and Clinical Evidence}

\begin{figure}[!htb]
    \centering
    \includegraphics[width=\columnwidth]{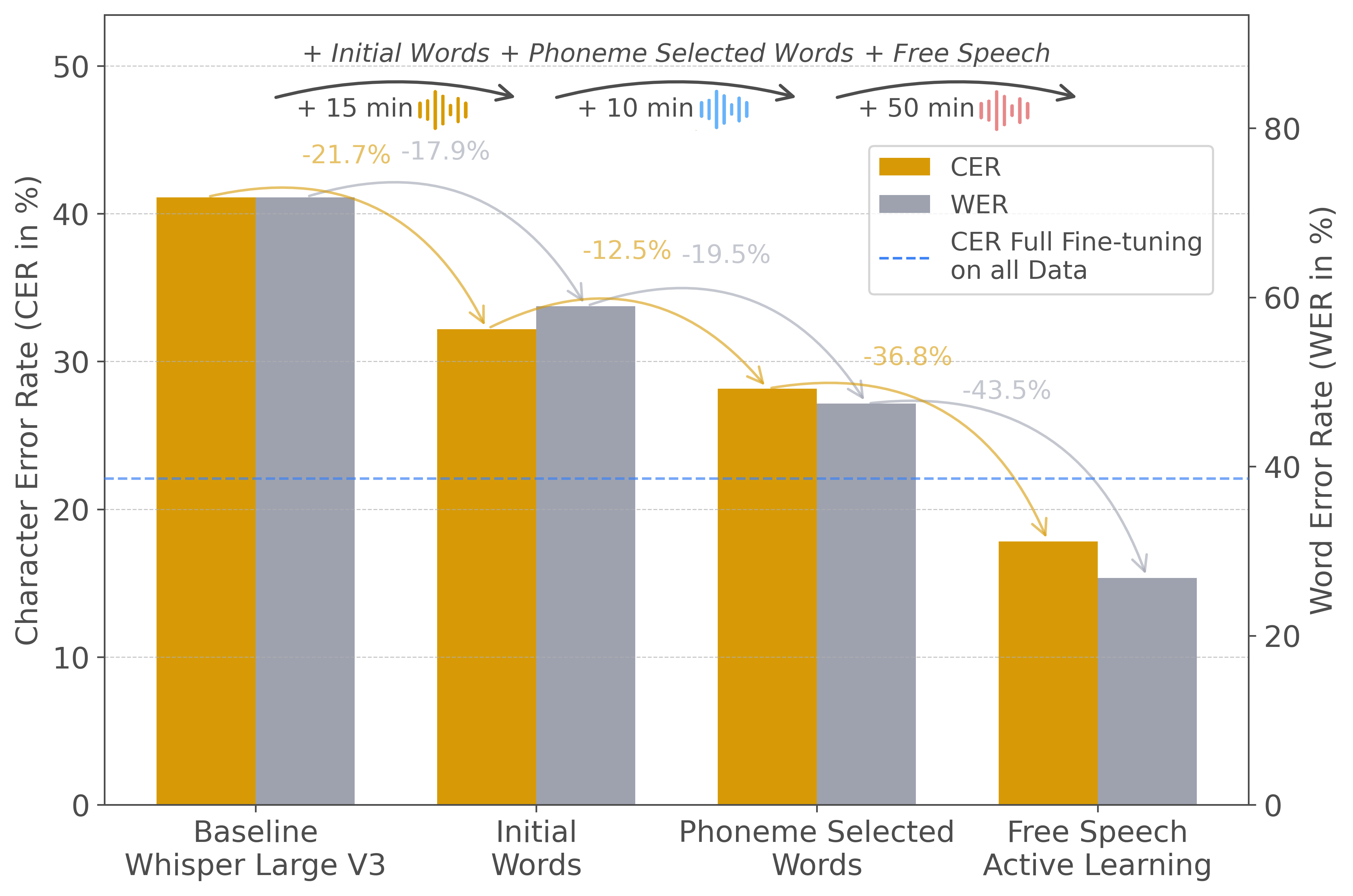}
    \caption{Longitudinal user study with a teenage child with structural speech impairment. The system reduces word error rate (WER) from a non-functional $\sim$70\% to usable $\sim$25\% within under 90 minutes of total interaction. Uncertainty-aware active learning (orange/grey bars) eventually outperforms the full fine-tuning baseline (blue dashed line), demonstrating the effectiveness of VI-LoRA in data-scarce regimes.}
    \label{fig:Results}
\end{figure}

We deployed \textit{Adapt4Me} in a home setting for a teenage user with structural speech impairment. As shown in Figure~\ref{fig:Results}, only 75 minutes of cumulative interaction (recording and correction) reduced the Word Error Rate (WER) from a non-functional 70\% of a Whisper Large baseline \cite{radford2023robust} to a usable level of approximately 25\%. The proposed active learning pipeline with VI-LoRA fine-tuning outperformed a brute-force full-parameter fine-tuning baseline (blue dashed line) while using substantially less data, demonstrating the efficiency of targeting high-uncertainty phonemes.
Qualitative analysis further shows that personalization restores communicative intent. Rather than producing semantic hallucinations, the adapted model makes phonetically plausible errors that remain intelligible to humans. For example, when dictating Swiss train stations (\textit{``Wiedikon, Enge, Thalwil, Baar''}), the baseline hallucinated unrelated sentences, whereas \textit{Adapt4Me} produced \textit{``Vidikon, Enne, Talwil, Borg''}. Despite high WER, these phonetic spellings are understandable in context, making the system useful.
Prior work \cite{pokel2025phonemes} validated model-predicted phoneme difficulties against clinical logopedic assessments. The strong correlation between the model's internal uncertainty estimates and therapists' evaluations confirms that the visual feedback provided by \textit{Adapt4Me} is clinically meaningful rather than a mere statistical artifact.

\section{Future Work}

\textit{Adapt4Me} offers a novel platform for large-scale speech science. By aggregating anonymized model states, specifically the trajectories of uncertainty heatmaps, researchers could perform meta-analysis on speech patterns without manual annotation. This could reveal latent population-level clusters, such as distinguishing between age-related phonetic drift and pathology-driven changes, effectively turning the ASR model into a diagnostic sensor.

\begin{acks}

We would like to thank Dr. Corinne Mathys Zulauf for her logopedic consulting, Dr. Samra Hamzic for valuable discussion and Philipp Guldimann for the initial data collection and annotation.

\end{acks}


\bibliographystyle{ACM-Reference-Format}
\bibliography{sample-base}










\end{document}